\documentclass[aps,prd,a4paper, preprint, superscriptaddress,preprintnumbers,floatfix,nofootinbib,amsmath,amssymb,table]{revtex4-1}
\usepackage{comment}
\usepackage{graphicx}
\usepackage{amsmath, mathrsfs}
\usepackage{subfigure}
\usepackage[usenames]{color}
\usepackage{xcolor}
\usepackage{amssymb}
\usepackage{bbm}
\usepackage{csquotes}
\usepackage{booktabs}
\usepackage{tikz}
\usepackage{soul}
\usetikzlibrary{positioning, decorations.pathmorphing}
\usepackage{xcolor}
\usepackage{epstopdf}
\newcommand{\bea}{\begin{aligned}}
\newcommand{\eea}{\end{aligned}}
\newcommand{\beq}{\begin{equation}}
\newcommand{\eeq}{\end{equation}}

\newcommand{\bse}{\begin{subequations}}
\newcommand{\ese}{\end{subequations}}

\usepackage{hyperref}
\hypersetup{
     colorlinks   = true,
     citecolor    = red,
     linkcolor    = blue,
     urlcolor     = blue,
}

\newcommand{\bmm}{\begin{multline}}
\newcommand{\emm}{\end{multline}}

\begin{document}
\title{Neutrino Fluence influenced by Memory Burdened Primordial Black Holes}
\author{Arnab Chaudhuri}
\email{arnab.chaudhuri@nao.ac.jp}
\email{arnab.chaudhuri@vit.ac.in}
\affiliation{Division of Science, National Astronomical Observatory of Japan, Mitaka, Tokyo 181-8588, Japan.}
\affiliation{Department of Physics, School of Advanced Sciences, Vellore Institute of Technology, Vellore-632014, Tamil Nadu, India.}
\author{Koushik Pal}
\email{palkoushik01@gmail.com}
\affiliation{School of Physics,  University of Hyderabad, Hyderabad - 500046,  India}
\author{Rukmani Mohanta}
\email{rmsp@uohyd.ac.in}
\affiliation{School of Physics,  University of Hyderabad, Hyderabad - 500046,  India}

\begin{abstract}
We study the impact of quantum gravitational memory burden—a backreaction effect that suppresses black hole evaporation—on neutrino signals from primordial black holes (PBHs). This suppression, modeled via a parameter \( k \), reduces the high-energy muon neutrino fluence, particularly during the late stages of evaporation. We also consider beyond-the-Standard-Model scenarios in which heavy neutral leptons (HNLs) are emitted by PBHs and subsequently decay, injecting secondary neutrinos that partially mitigate the suppression in the MeV–GeV range. We compute the full time-integrated neutrino spectrum and evaluate the expected IceCube event rates across the \((k, m_N)\) parameter space. We analyze both single-source burst scenarios and the cumulative Galactic contribution assuming PBHs trace a realistic dark matter halo distribution. Even under optimistic proximity assumptions, the predicted event rates remain far below IceCube sensitivity, and population-level stacking within current observational bounds on the PBH abundance does not yield an observable signal in the considered mass range for PBH abundances consistent with existing observational constraints. These results demonstrate that entropy-suppressed evaporation substantially weakens neutrino detectability of light PBHs and must be consistently incorporated in future multi-messenger searches.
\end{abstract}

\maketitle

\section{Introduction}
\label{introduction}

The Standard Model (SM) successfully explains a wide range of particle physics phenomena, yet it fails to account for dark matter~\cite{Planck:2018vyg}, neutrino masses, the matter–antimatter asymmetry, and cosmic acceleration. These gaps motivate the search for physics beyond the SM (BSM), often realized through scalar extensions~\cite{Chaudhuri:2022sis, Morrissey:2012db, Carena:2019une, Chiang:2019oms, Kang:2017mkl, Karmakar:2023ixo, Borah:2023zsb, Kikuta:2014eja, Chaudhuri:2024vrd,Chaudhuri:2026ovn,Chaudhuri:2025cjp,Chaudhuri:2025ybh,Srivastava:2025oer,Das:2026zuo}. The early universe offers a promising setting to explore BSM signatures, particularly through the physics of primordial black holes (PBHs).

PBHs could have formed from large density fluctuations in the early universe~\cite{Escriva:2022duf, Zeldovich:1967lct, Carr:1974nx,Hawking:1987bn,Polnarev:1988dh,Hawking:1982ga,Kodama:1982sf,Page:1976ki,Deng:2017uwc,Liu:2019lul,Jung:2021mku,Baker:2021nyl, Hayward:1989jq,Page:2004xp,Page:1976df,MacGibbon:2007yq,Page:1977um}, and their Hawking evaporation~\cite{Page:2004xp} provides a direct probe of high-energy physics in the pre-BBN epoch. The formation time and corresponding mass of PBHs are related through $M \simeq 10^{15}~\mathrm{g}~(t / 10^{-23}~\mathrm{s})$, such that PBHs forming at the Planck time ($t \sim 10^{-43}$ s) would have masses near $10^{-5}$ g, while those forming at $t \sim 1$ s could reach $\sim 10^5 M_\odot$~\cite{Carr:2020xqk}.

In the semiclassical approximation, PBHs with masses near $5 \times 10^{14}$ g are completing their evaporation today, having lifetimes comparable to the age of the Universe and ~\cite{Page:1976df}. PBHs with masses below $10^9$~g would have evaporated before nucleosynthesis, but their decays may still leave observable relics, such as entropy injection, dark matter, and high-energy neutrinos~\cite{Schumann:2019eaa,Billard:2021uyg,Bird:2022wvk, Green:2024bam,Carr:2020xqk,Chaudhuri:2020wjo, Chattopadhyay:2022fwa,Chaudhuri:2023aiv,Chaudhuri:2026zyx}.

Recent gravitational wave detections~\cite{LIGOScientific:2016aoc, LIGOScientific:2020kqk} have renewed interest in PBHs across a wide mass range. Their evaporation products are constrained by gamma-ray~\cite{Laha:2020ivk, Ballesteros:2019exr, Arbey:2019mbc, Wright:1995bi, Page:1976wx}, charged particle~\cite{Dasgupta:2019cae, DeRocco:2019fjq, Laha:2019ssq, CMS:2019uhm}, and neutrino~\cite{Bernal:2022swt, DeRomeri:2021xgy, Wang:2020uvi, Dasgupta:2019cae} measurements, as well as by BBN and CMB observations~\cite{Carr:2020gox, Carr:2009jm}.

Evaporating PBHs emit all particles lighter than their instantaneous Hawking temperature. Among these, neutrinos provide a key detection channel, especially when enhanced by BSM physics such as heavy neutral leptons (HNLs)~\cite{Chen:2023tzd, Bernal:2022pue}. HNLs can emerge in seesaw-like models and play roles in leptogenesis~\cite{Borah:2022uos}, and may be thermally produced or emitted directly by PBHs. In particular, decay-sourced neutrinos can arise from secondary decays of SM particles or from exotic processes such as HNL decays.

In this work, we incorporate the memory burden effect~\cite{Dvali:2018xpy, Dvali:2020wft}, which arises from the breakdown of semiclassical approximations as a black hole loses a significant fraction of its mass. This effect, explored in several recent studies~\cite{Barman:2024ufm, Barman:2024kfj, Thoss:2024hsr, Basumatary:2024uwo, Chianese:2024rsn, Chianese:2025wrk, Athron:2024fcj, Calabrese:2025sfh, Dvali:2024hsb, Zantedeschi:2024ram, Datta:2023vbs, Bandyopadhyay:2025ast, Chaudhuri:2025asm}, suppresses late-time particle emission and modifies the black hole’s evaporation profile.

We systematically analyze how the memory burden modifies the evaporation dynamics, the instantaneous temperature evolution, and the resulting particle spectra, particularly for neutrinos. We study how memory-burdened PBHs, if located within the local parsec-scale neighborhood, could produce detectable neutrino bursts at IceCube~\cite{Coogan:2020tuf, Carr:2016drx}. In addition, we assess how HNL production affects both neutrino and photon spectra.

This paper presents a detailed calculation of the time-integrated muon neutrino flux at IceCube, comparing SM-only and HNL-enhanced scenarios under varying memory burden strengths~\cite{IceCube:2025kve}. We assess how HNL production affects the resulting neutrino spectra and detection prospects at IceCube. Our analysis shows that the memory burden can significantly suppress neutrino fluence, and that detection prospects are strongly constrained by distance and particle physics assumptions.

\section{Theory of PBHs and Memory Burden}
\label{BH}

\subsection{PBHs and Neutrino Production}

Evaluated on comoving (uniform-Hubble) slices, numerical simulations indicate that gravitational collapse into a primordial black hole (PBH) occurs when the density contrast exceeds a critical threshold, $\delta \gtrsim \delta_c$, with $\delta_c \simeq 0.4$ \cite{Musco:2004ak,Harada:2013epa,Escriva:2022duf,Ozsoy:2023ryl}, depending mildly on the perturbation profile and the equation of state. Other PBH formation mechanisms have also been proposed in the literature, including the collapse of cosmic defects such as cosmic strings or domain walls \cite{Hawking:1987bn,Polnarev:1988dh}, as well as bubble collisions during first-order phase transitions in the early Universe \cite{Hawking:1982ga,Kodama:1982sf}.

Since the subsequent analysis is insensitive to the detailed modeling of PBH formation and to the precise value of $\delta_c$, we do not pursue PBH formation mechanisms further. Instead, we focus on the evaporation properties and observational signatures of an assumed PBH population.

The PBH mass depends on the epoch of formation, as summarized in Table~\ref{tab:pbh_mass_time} \cite{Carr:2020gox}. This dependence arises from the causal horizon mass at formation, typically given by
\begin{equation}
M \simeq 10^{15}~\text{g} \left(\frac{t}{10^{-23}~\text{s}}\right).
\end{equation}

\begin{table}[h!]
\centering
\begin{tabular}{|l|l|}
\hline
Time after the Big Bang(Seconds) & Typical PBH Mass(g) \\
\hline
$10^{-23}$ & $10^{15}$ \\
$10^{-21}$ & $10^{17}$ \\
$10^{-19}$ & $10^{19}$ \\
$10^{-17}$ & $10^{21}$ \\
$10^{-15}$ & $10^{23}$ \\
\hline
\end{tabular}
\caption{Primordial black hole mass as a function of formation time, computed using the horizon-mass relation in Eq.~(2) during radiation domination.}
\label{tab:pbh_mass_time}
\end{table}

PBHs radiate thermally via Hawking evaporation \cite{Hawking:1974rv, Hawking:1975vcx}, emitting all species with masses below the Hawking temperature,
\begin{equation}
T_{\mathrm{H}} = \frac{1}{8 \pi G M} \sim 1~\text{TeV} \left( \frac{10^{10}~\text{g}}{M} \right).
\end{equation}
In the semi-classical limit, Schwarzschild PBHs (negligible charge/spin) lose mass at the rate
\begin{equation}
\frac{dM}{dt} = -\varepsilon(M) \frac{M_p^4}{M^2},
\end{equation}
where $M$ is the PBH mass, $M_p \equiv (8\pi G)^{-1/2}$ is the reduced Planck mass, and $\varepsilon(M)$ is a dimensionless emissivity factor encoding the total number of particle degrees of freedom emitted by the black hole. In general, $\varepsilon$ depends weakly on $M$ through the opening of additional particle emission channels as the Hawking temperature increases~\cite{Cheek:2022mmy, Cheek:2021odj, MacGibbon:1990zk, MacGibbon:1991tj, Halzen:1995hu}.  this work, we treat $\varepsilon$ as approximately constant over the mass range of interest.

 The emissivity factor can be written as a sum over particle species,

\begin{equation}
\varepsilon(M) = \sum_{i = \text{dofs}} \frac{g_i}{128 \pi^3} \int_0^{\infty} \frac{x \, \mathcal{G}_{s_i}(x)}{\exp(x) - (-1)^{2s_i}} \, dx,
\end{equation}

 The index $i$ runs over all particle degrees of freedom such as standard Model particles, possible BSM states and each helicity, color, particle/antiparticle counted separately. \( g_i \) is the internal degrees of freedom of a specific species \( i \) such as spin multiplicity, color, and Particle vs antiparticle. The value of $g_i$ in case of photons, Dirac fermions and gluons are 2, 4 and 16 respectively. \( \mathcal{G}_{s_i}(x) \) denotes the absorption probability—often referred to as the greybody factor—which accounts for the partial backscattering of emitted particles due to centrifugal and/or gravitational potentials induced by the curved spacetime surrounding the black hole. These quantities are typically evaluated numerically, following the methods described in Ref. \cite{Page:1976ki}.

In our numerical calculations, we use \texttt{BlackHawk v2.0}~\cite{Arbey:2019mbc, Arbey:2021mbl}, which includes accurate greybody corrections and secondary decays. Assuming constant \( \varepsilon \), the PBH mass evolves as
\begin{equation}
M(t) = M_{\text{in}} \left( 1 - \frac{t}{\tau_d} \right)^{1/3},
\end{equation}
where
\begin{equation}
\tau_d = \frac{M_{\text{in}}^3}{3 \varepsilon M_p^4} \sim 428~\text{s} \left( \frac{4.07 \times 10^{-3}}{\varepsilon} \right) \left( \frac{M_{\text{in}}}{10^{10}~\text{g}} \right)^3.
\label{lifetime}
\end{equation}
The value \( \varepsilon \sim 4.07 \times 10^{-3} \) \cite{MacGibbon:1991tj} corresponds to the SM particle content including all relativistic species and accounts for greybody suppression. Unless stated otherwise, all numerical results are obtained assuming
a monochromatic primordial black hole mass
$M_{\rm PBH} = 10^{10}\,\mathrm{g}$ at formation.
Such PBHs complete their evaporation on timescales
$\mathcal{O}(10^2)$--$\mathcal{O}(10^3)\,\mathrm{s}$,
making them relevant for burst-like neutrino emission.
The time-dependent Hawking temperature $T_H(M_{\rm PBH})$ and
the corresponding greybody factors are evaluated self-consistently
throughout the evaporation history, including the memory-burden regime. We have verified that varying the initial PBH mass within
$10^9$--$10^{11}\,\mathrm{g}$ does not qualitatively affect
the conclusions regarding memory-burden suppression.

The instantaneous emission rate for particle species \( i \) is
\begin{equation}
\frac{d^2 N_i}{dE \, dt}(M) = \frac{g_i}{2\pi} \, \frac{\mathcal{G}_{s_i}(E)}{e^{8\pi G M E} - (-1)^{2s_i}},
\end{equation}
and the total fluence integrated over the lifetime \( \tau \) is
\begin{equation}
\frac{dN_i}{dE} = \int_0^{\tau} dt \, \frac{d^2N_i}{dE \, dt}\big(M(t)\big),
\end{equation}
including both directly Hawking-emitted and secondary (decay-sourced) contributions. The latter includes neutrinos from decays of SM particles and from exotic sources such as HNLs. In all numerical calculations, the time-integrated neutrino spectrum
is obtained by integrating the instantaneous emission rate
over the full PBH lifetime,
\begin{equation}
\frac{dN_\nu}{dE} = \int_0^{\tau_{\rm PBH}} dt\,
\frac{d^2N_\nu}{dE\,dt},
\end{equation}
where $\tau_{\rm PBH}$ denotes the total evaporation time,
including the modification due to the memory-burden effect
discussed in Sec. \ref{mb}.

\subsection{Memory Burden Effects}
\label{mb}
The loss of PBH mass can also be expressed as~\cite{MacGibbon:1990zk, Page:1976wx, Drees:2015exa, Mazde:2022sdx}
\begin{equation}
\frac{\mathrm{d}M_{\mathrm{PBH}}}{\mathrm{d}t} = 
- \frac{\mathcal{G} \, g_{\mathrm{SM}}}{30720 \pi G^2 M_{\mathrm{PBH}}^2},
\label{decay}
\end{equation}
where \( \mathcal{G} \sim 3.8 \) is a numerical factor obtained from the greybody-integrated emission spectra, and \( g_{\mathrm{SM}} \sim 106.75 \) denotes the total number of relativistic degrees of freedom in the Standard Model.

Quantum gravitational effects may significantly slow down the evaporation process through the "memory burden" mechanism, which becomes relevant when the black hole retains quantum information about its earlier states~\cite{Dvali:2018xpy, Dvali:2020wft}. This effect modifies the dynamics of evaporation, particularly during the late stages of the PBH lifetime.

The onset of the memory burden occurs at  \( t_q = \tau_{\mathrm{PBH}} (1 - q^3) \), where \( q \in (0,1) \) parameterizes the fraction of the initial PBH mass that remains when quantum-gravitational suppression sets in. In this work, we adopt \( q = 1/2 \) as a representative benchmark value, following~\cite{Chianese:2024rsn, Dvali:2020wft}. This corresponds to the burden becoming relevant when the PBH has lost 50\% of its original mass. At this point, approximately 87.5\% of the evaporation lifetime has elapsed. This choice balances early and late onset scenarios: larger \( q \) values delay suppression (reducing its observational impact), while smaller \( q \) lead to earlier onset and stronger suppression. Our results show that while the precise value of \( q \) mildly affects the spectral shape, the dominant influence on the observable neutrino fluence is governed by the entropy-suppression exponent \( k \).
The remaining PBH mass at the onset of memory-burden regime is
\begin{equation}
M_{\mathrm{PBH}}^{\mathrm{mb}} = q M_{\mathrm{PBH}}.
\end{equation}

In the burdened regime, the mass loss rate is suppressed by the black hole entropy:
\begin{equation}
\frac{\mathrm{d}M_{\mathrm{PBH}}^{\mathrm{mb}}}{\mathrm{d}t} =
\frac{1}{S(M_{\mathrm{PBH}})^k} \frac{\mathrm{d}M_{\mathrm{PBH}}}{\mathrm{d}t}, \quad S = 4\pi G M^2, \quad k > 0,
\label{rate}
\end{equation}
where \( k \) is a positive model-dependent parameter that quantifies the strength of suppression. Higher values of \( k \) yield stronger suppression of late-time emission.

Integrating the modified mass loss rate yields the time evolution of the PBH mass in the memory-burdened regime~\cite{Chianese:2025wrk}:
\begin{equation}
M_{\mathrm{PBH}}^{\mathrm{mb}}(t) = M_{\mathrm{PBH}} \left[ 1 - \Gamma_{\mathrm{PBH}}^{(k)} (t - t_q) \right]^{\frac{1}{3+2k}},
\end{equation}
where
\begin{equation}
\Gamma_{\mathrm{PBH}}^{(k)}=\frac{\mathcal{G}{g_\mathrm{SM}}}{7680 \pi} 2^{k}(3+2k)M_{p} \left(\frac{M_{p}}{M^\mathrm{mb}_{\mathrm{PBH}}}\right)^{3+2k}.
\end{equation}
Note that $\mathcal{G}$ is the numerical factor obtained from greybody-integrated emission spectra and $g_{\rm {SM}}$ is the SM relativistic degrees of freedom. 

The total PBH lifetime in the presence of memory burden effects is then given by
\begin{equation}
\tau_{\mathrm{PBH}}^{(k)} = t_q + \left(\Gamma_{\mathrm{PBH}}^{(k)}\right)^{-1} 
\simeq \left(\Gamma_{\mathrm{PBH}}^{(k)}\right)^{-1},
\end{equation}
where the approximation follows from the fact that the memory-burdened phase dominates the late-time evolution.

The formalism in this section closely follows \cite{Chianese:2025wrk}, originally developed for gamma-ray emission, and is here adapted to neutrino production with the inclusion of flavor oscillations and heavy neutral leptons. The neutrino emission spectrum from the former standard semiclassical Hawking treatment can be read as:
\begin{equation}
\frac{\mathrm{d}^2 N_{\nu}}{\mathrm{d}E \mathrm{d}t} = 
\frac{g_{\nu}}{2\pi} 
\frac{\mathcal{F}(E, M_{\mathrm{PBH}})}{e^{E/T_{\mathrm{H}}} + 1},
\end{equation}
where \( \mathcal{F}(E, M_{\mathrm{PBH}}) \) includes greybody effects, and \( g_\nu \) counts neutrino degrees of freedom.

The corresponding neutrino emission spectrum is suppressed relative to the standard semiclassical Hawking emission and can be expressed as:
\begin{equation}
\frac{\mathrm{d}^2 N_{\nu}^{\rm mb}}{\mathrm{d}E \mathrm{d}t}=S(M_{\mathrm{PBH}})^{-k} \frac{\mathrm{d^2}N_{\nu}}{\mathrm{d}E \mathrm{d}t},
\end{equation}

As a consequence, high-energy neutrino signals originating from burdened survivor PBHs with masses \( M_{\text{PBH}} \lesssim 10^9\,\text{g} \) can be significantly suppressed compared to the standard evaporation scenario.

The total all-flavor differential flux from Galactic and extragalactic populations is given by
\begin{equation}
\frac{\mathrm{d}^2\Phi_{\nu}}{\mathrm{d}E\mathrm{d}\Omega}=\sum_{\alpha} \left( \frac{ \mathrm{d}^2 \phi^{\mathrm{gal}}_{\nu_{\alpha}} }{ \mathrm{d}E \, \mathrm{d}\Omega } + \frac{ \mathrm{d}^2 \phi^{\mathrm{egal}}_{\nu_{\alpha}} }{ \mathrm{d}E \, \mathrm{d}\Omega } \right),
\end{equation}
where the two terms account for the line-of-sight integration over Galactic dark matter halo profiles and for the cosmological redshifted flux from uniformly distributed extragalactic PBHs, respectively. These are computed using standard cosmological inputs and PBH mass spectra, as detailed in Section~\ref{res}.

\section{Results}
\label{res}

Here we present the total time-integrated muon neutrino flux (or neutrino fluence) from PBH evaporation, incorporating both the standard Hawking phase and the quantum backreaction from the memory burden (MB) effect. We consider two scenarios: (i) evaporation into Standard Model (SM) particles only, and (ii) SM extended by a heavy neutral lepton (HNL) with masses $m_N = 0.2$~GeV and $1.0$~GeV. The HNLs are assumed to be kinematically accessible during PBH evaporation, with decay widths into neutrino-rich final states as computed using \texttt{BlackHawk v2.0} secondary modules. For each case, we study how varying the MB suppression parameter $k$ affects the neutrino fluence over an observation window of $\tau = 100$~s.

The spectra are computed using a modified \texttt{BlackHawk v2.0} setup, in which the evaporation rate is entropy-suppressed by a factor $S^{-k}$ in the final stages, and the decay-sourced neutrinos from both SM and HNL final states are fully included. The emission is assumed to be flavor-diagonal at the source.

\begin{figure}[h]
  \centering
  \begin{minipage}[b]{0.47\textwidth}
    \includegraphics[width=\textwidth]{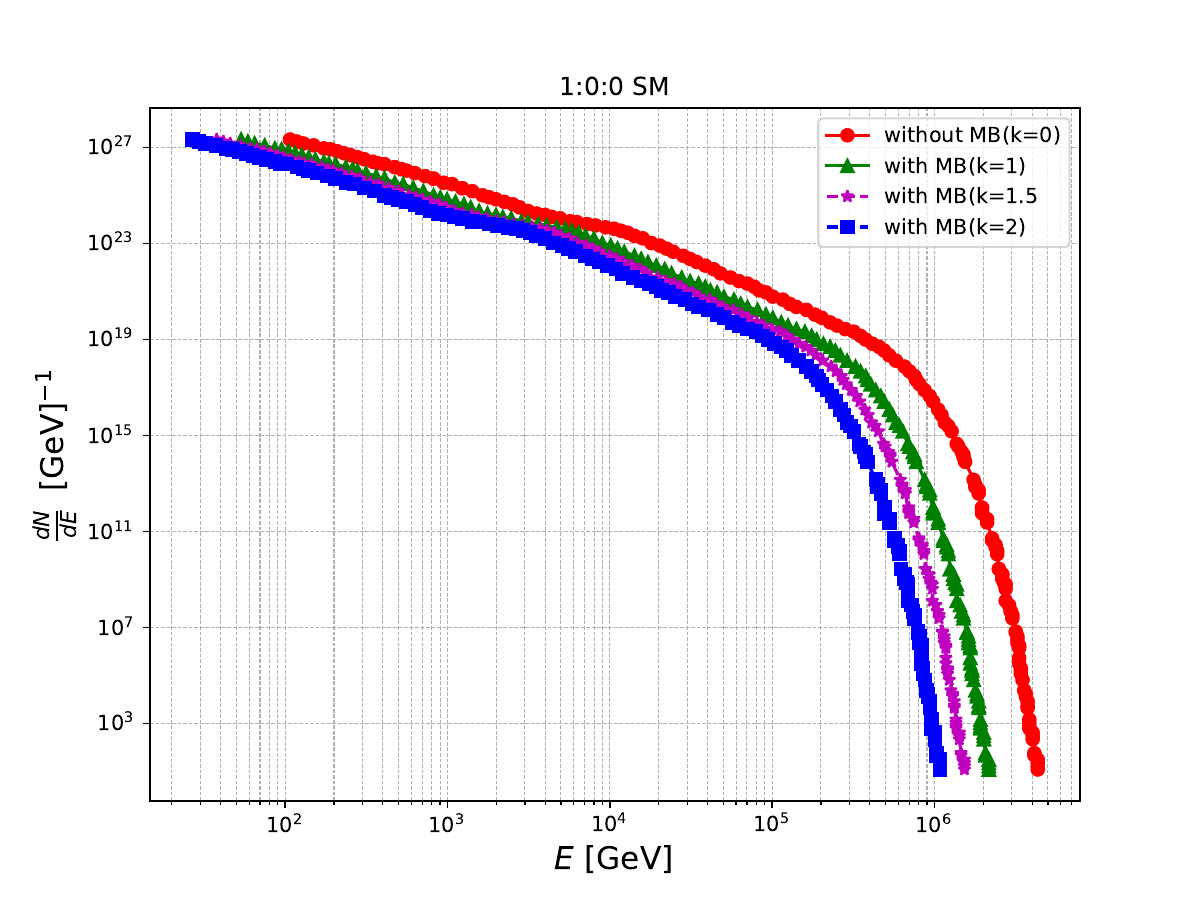}  
  \end{minipage}
  \hspace*{.1cm}
  \begin{minipage}[b]{0.47\textwidth}
    \includegraphics[width=\textwidth]{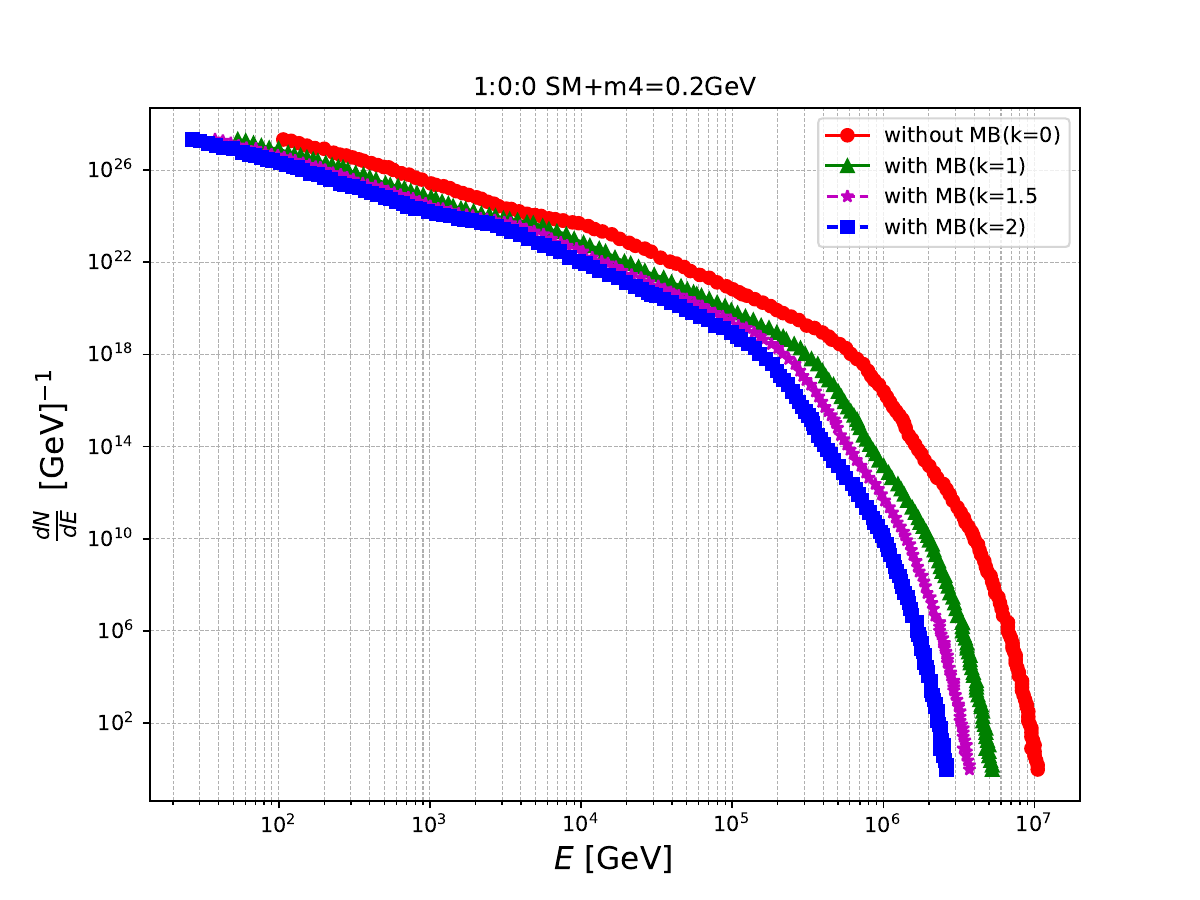}
  \end{minipage}
  \hspace*{.1cm}
  \begin{minipage}[b]{0.47\textwidth}
    \includegraphics[width=\textwidth]{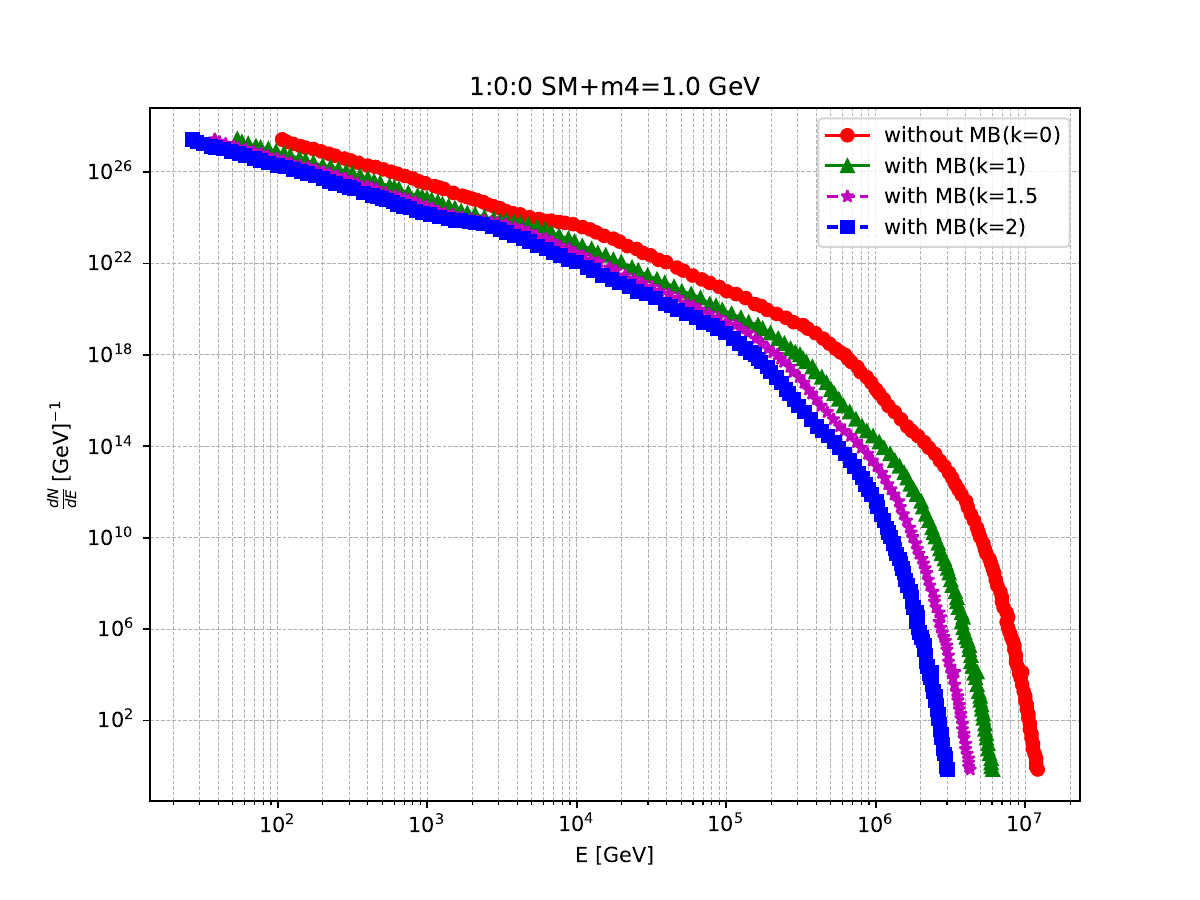}
  \end{minipage}
  \caption{Total time-integrated muon neutrino flux at Earth for an initial
$(1:0:0)$ flavor composition at the source, including neutrino oscillations.}
  \label{fig1}
\end{figure}

Figure~\ref{fig1} (top left) shows the total time-integrated muon neutrino flux at Earth,
after neutrino propagation and flavor conversion.
We assume an initial flavor composition
$(\nu_e : \nu_\mu : \nu_\tau) = (1:0:0)$ at the source, which is transformed
during propagation according to the Eq.(\ref{eq:Nevents}). The red curve represents the standard Hawking evaporation without memory burden ($k = 0$), while the green, magenta, and blue curves correspond to $k = 1$, $1.5$, and $2$, respectively.

We observe a clear suppression of the high-energy tail of the spectrum with increasing $k$. For example, at $E_\nu \sim 10^5$~GeV, the flux decreases by approximately an order of magnitude when comparing $k = 0$ and $k = 2$. This reflects the suppression of high-temperature emission near the final stages of PBH evaporation, as the memory burden slows down mass loss and reduces the Hawking temperature. The spectrum thus softens and peaks at lower energies.

Next, we incorporate HNLs into the PBH decay spectrum. In the top right and bottom panels of Figure~\ref{fig1}, we present the neutrino fluence for HNL masses $m_N = 0.2$~GeV and $1.0$~GeV, respectively. In both cases, the red curve denotes the $k = 0$ result, while other colors indicate increasing MB suppression.

The inclusion of HNLs enhances the overall neutrino fluence across the entire energy range. This enhancement is due to their decay into SM leptons and neutrinos, which inject secondary neutrinos after their production in PBH evaporation. The effect is stronger for larger HNL masses, since more energetic decay products populate the high-energy tail of the spectrum. 

Quantitatively, for $m_N = 1.0$~GeV and $k = 1.5$, the fluence at $E_\nu \sim 10^3$~GeV is enhanced by a factor of $\sim 3$ compared to the SM-only case with the same $k$ value. This compensating effect makes HNLs a compelling BSM addition for indirect detection strategies.

\begin{figure}[h]
  \centering
  \begin{minipage}[b]{0.47\textwidth}
    \includegraphics[width=\textwidth]{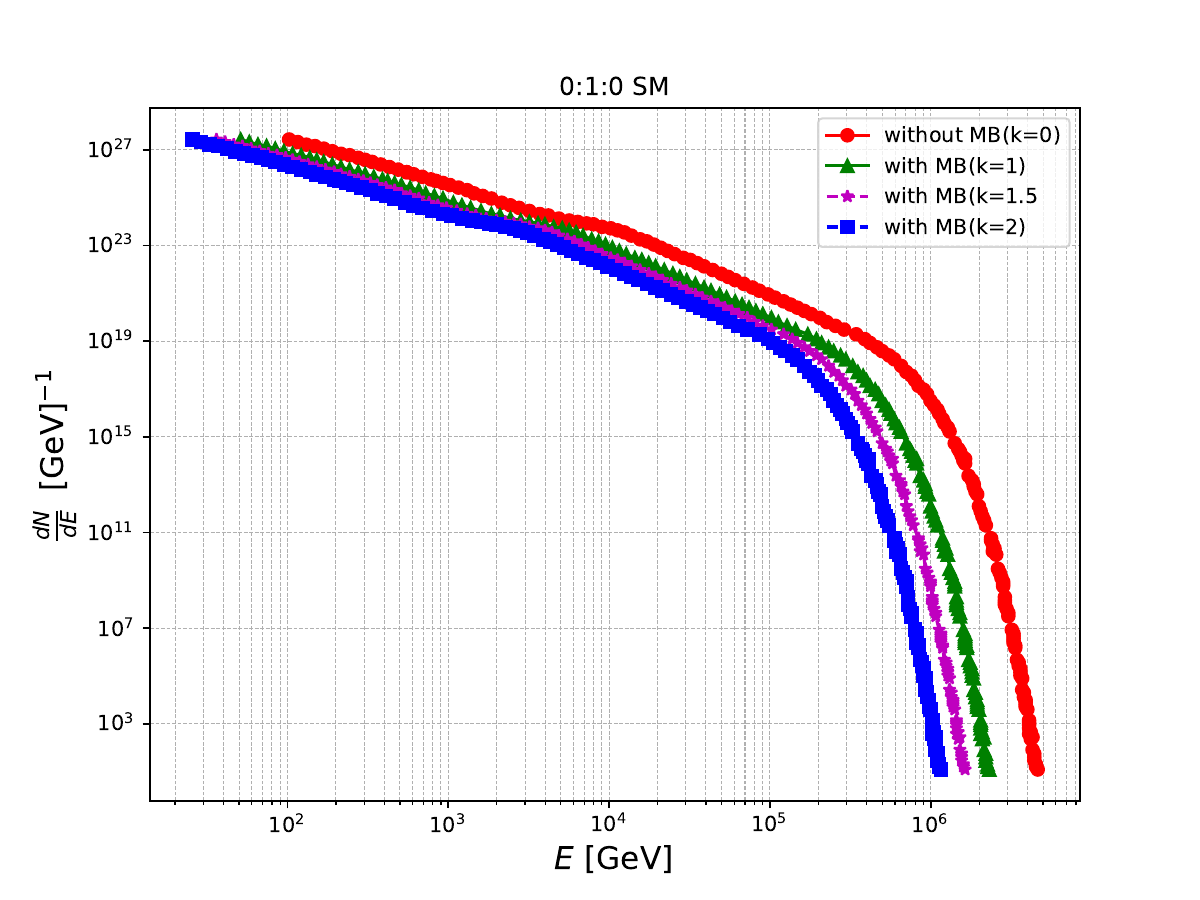}  
  \end{minipage}
  \hspace*{.1cm}
  \begin{minipage}[b]{0.47\textwidth}
    \includegraphics[width=\textwidth]{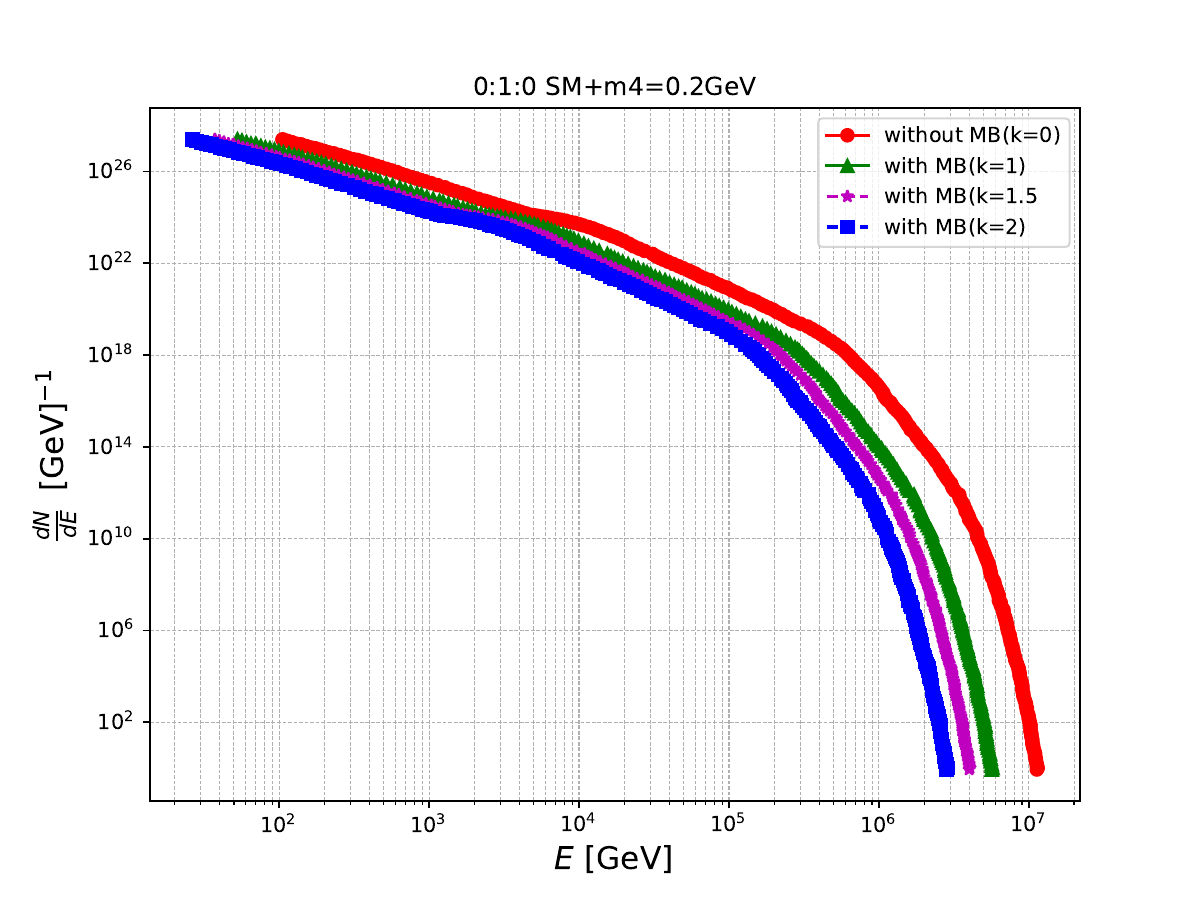}
  \end{minipage}
  \hspace*{.1cm}
  \begin{minipage}[b]{0.47\textwidth}
    \includegraphics[width=\textwidth]{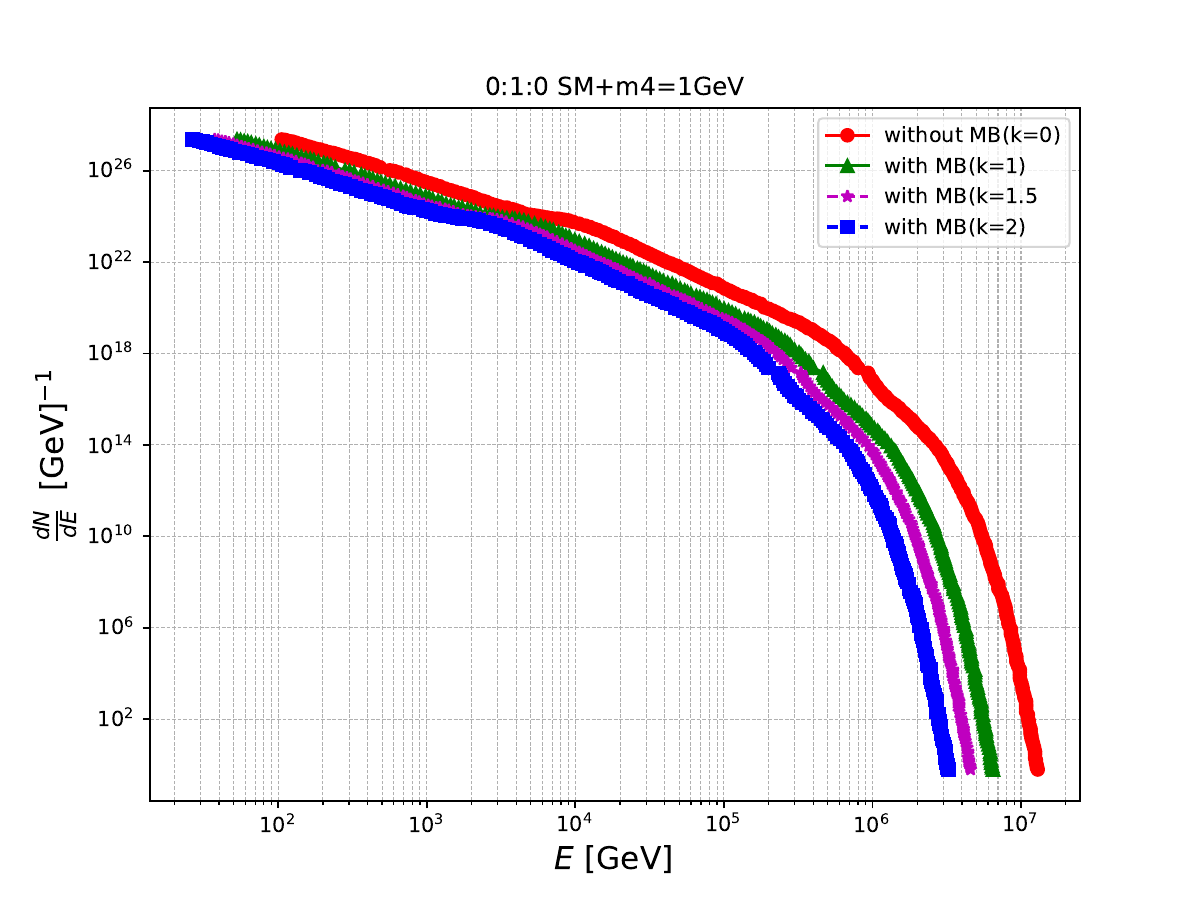}
  \end{minipage}
  \caption{Same as Fig.~\ref{fig1}, but for $0:1:0$ flavor composition at the source.}
  \label{fig2}
\end{figure}

\begin{figure}[h]
  \centering
  \begin{minipage}[b]{0.47\textwidth}
    \includegraphics[width=\textwidth]{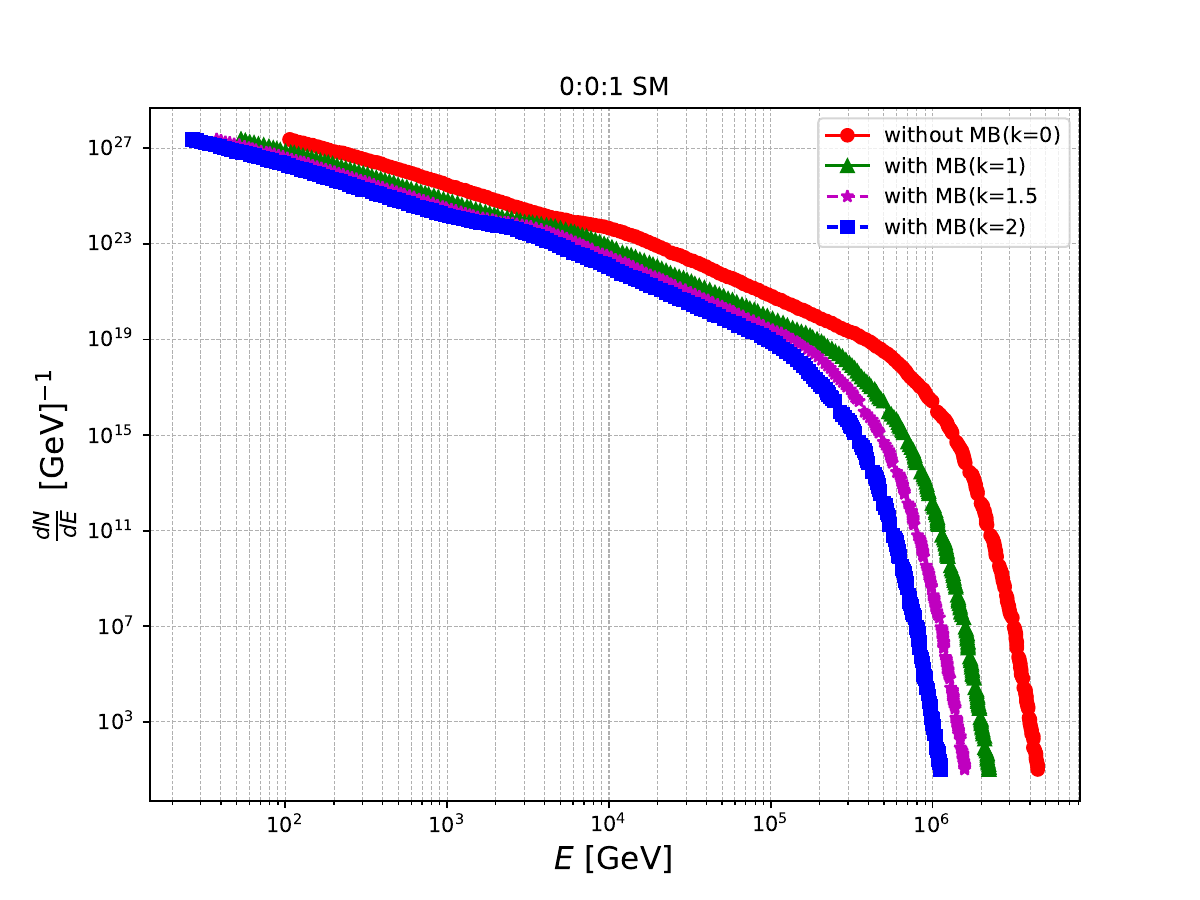}  
  \end{minipage}
  \hspace*{.1cm}
  \begin{minipage}[b]{0.47\textwidth}
    \includegraphics[width=\textwidth]{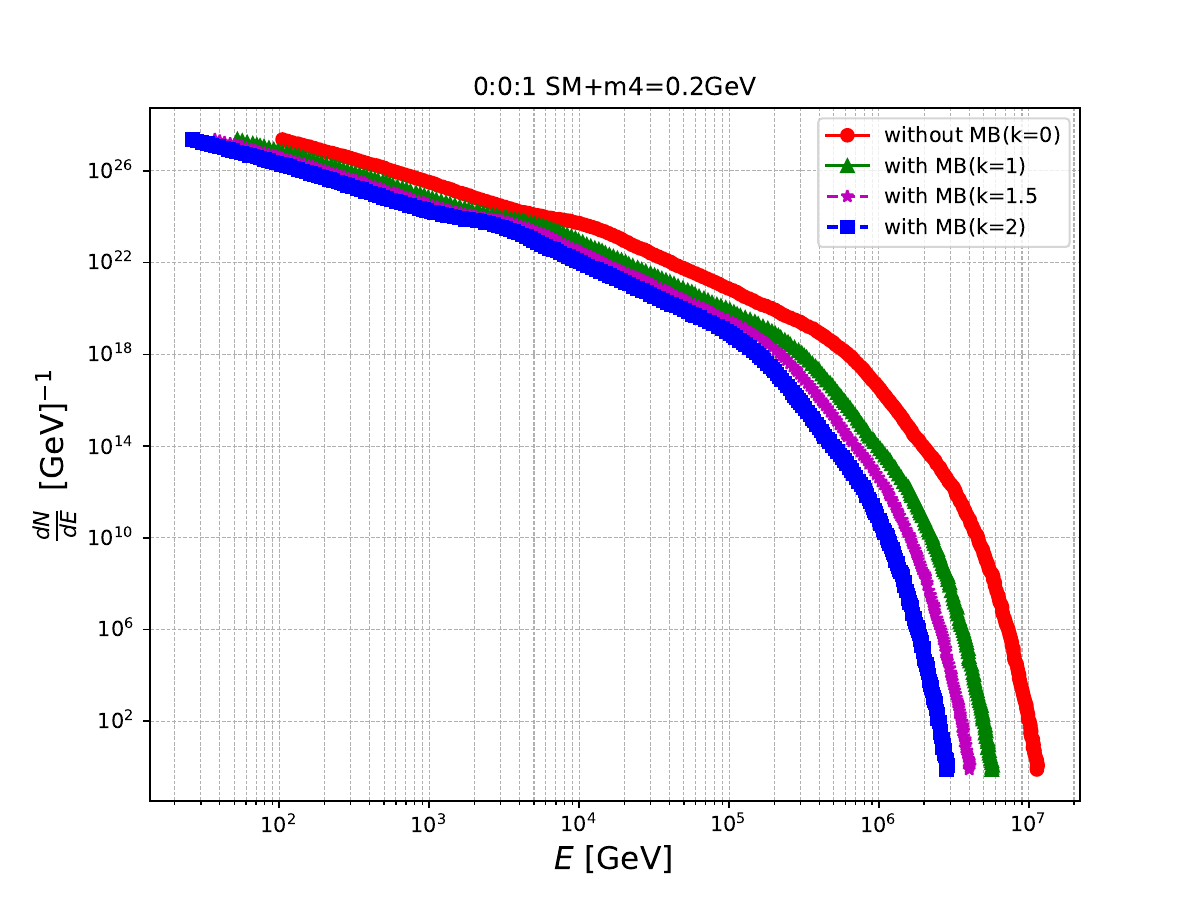}
  \end{minipage}
  \hspace*{.1cm}
  \begin{minipage}[b]{0.47\textwidth}
    \includegraphics[width=\textwidth]{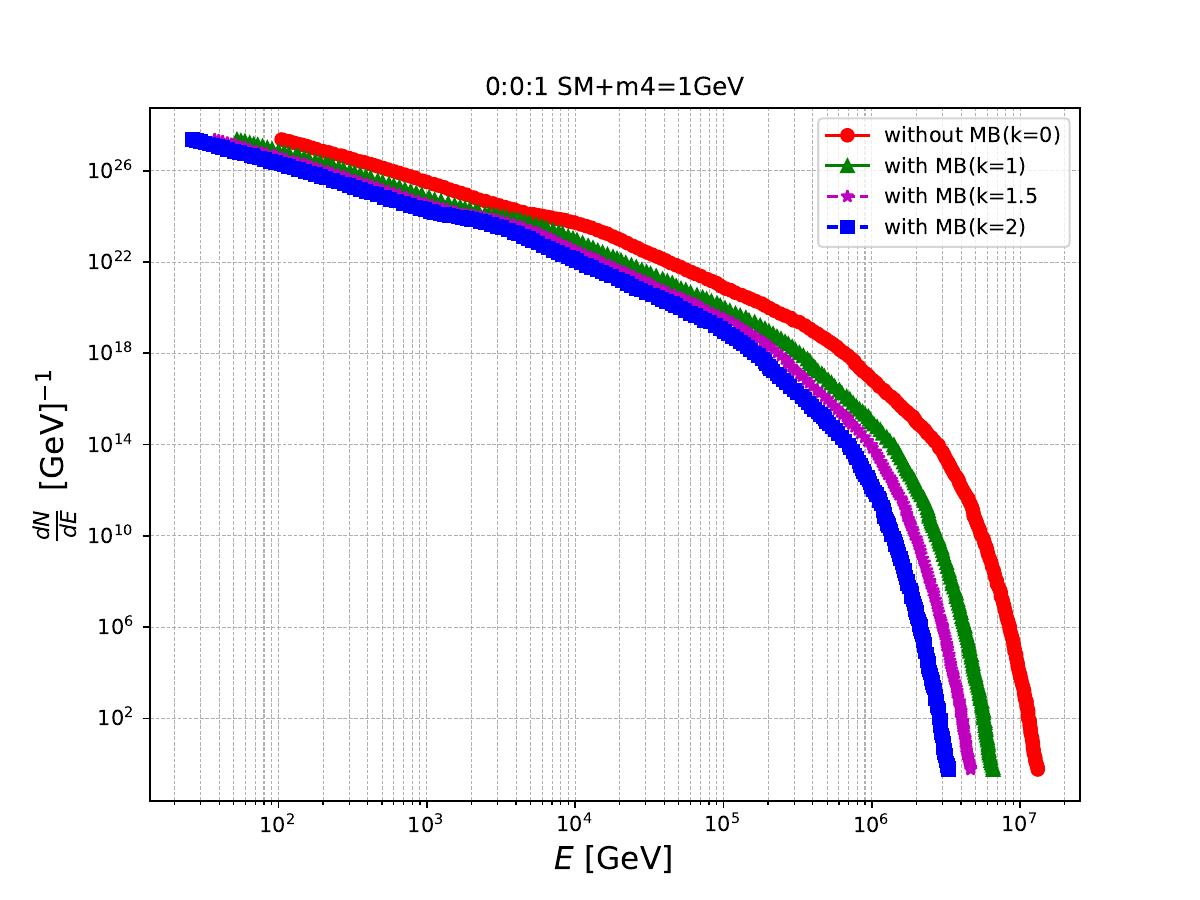}
  \end{minipage}
  \caption{Same as Fig.~\ref{fig1}, but for $0:0:1$ flavor composition at the source.}
  \label{fig3}
\end{figure}

Figures~\ref{fig2} and~\ref{fig3} show the fluences for $0:1:0$ and $0:0:1$ source compositions. The qualitative trends remain unchanged, confirming that memory burden suppression is flavor-blind at the source, and HNL decay contributions enhance all channels comparably.

In summary, the inclusion of the memory burden systematically reduces the high-energy neutrino fluence, especially for $k \gtrsim 1$. However, the addition of HNLs partially mitigates this suppression by opening new decay pathways that replenish the neutrino spectrum.

\begin{table}[h]
\centering
\begin{tabular}{|c|c|c|c|}
\hline
~~$k$~~ &~~ SM only~~ & SM + HNL ($m_N = 0.2$ GeV) & SM + HNL ($m_N = 1.0$ GeV) \\
\hline
0   & $1.2 \times 10^{11}$ & $1.5 \times 10^{11}$ & $2.1 \times 10^{11}$ \\
1.0 & $5.6 \times 10^{10}$ & $7.2 \times 10^{10}$ & $1.1 \times 10^{11}$ \\
2.0 & $1.1 \times 10^{10}$ & $2.8 \times 10^{10}$ & $4.3 \times 10^{10}$ \\
\hline
\end{tabular}
\caption{Representative values of the muon neutrino fluence $dN/dE$ (GeV$^{-1}$) at $E_\nu = 10^4$~GeV for different suppression factors $k$ and HNL masses using the full time-dependent emission spectrum integrated over the PBH lifetime.}
\label{tab:fluence_summary}
\end{table}
The fluence values in Table~\ref{tab:fluence_summary} highlight the competing roles of MB suppression and HNL-induced enhancement. The numbers are computed using the full time-dependent emission spectrum integrated over the PBH lifetime.

To assess the observational prospects at IceCube, we estimate the number of detectable muon neutrino events from a PBH explosion occurring at distance $D = 0.1$ pc. Using the fluence at $E_\nu = 10^4$ GeV for $m_N = 1.0$ GeV and $k = 2$ ($dN/dE \sim 1.1 \times 10^{10}~\text{GeV}^{-1}$), and assuming an effective area $A_{\text{eff}} \sim 1~\text{m}^2$, the number of events is:
\begin{equation} \label{eq:Nevents}
\frac{d^2\Phi_\nu}{dE\,d\Omega} \sim \frac{dN}{dE} \cdot \frac{A_{\text{eff}}}{4\pi D^2} \cdot \Delta E \sim 10^{-19}.
\end{equation}
The energy bin $\Delta E \sim 10^4~\text{GeV}$ is taken conservatively, but even under this estimate the signal is undetectable at parsec distances.
The effective area \( A_{\text{eff}} \sim 1~\text{m}^2 \) is a conservative estimate based on IceCube’s performance at neutrino energies around \( E_\nu \sim 10^4~\text{GeV} \), where the detector has moderate sensitivity to muon neutrinos. According to IceCube technical documentation and effective area curves~\cite{IceCube:2016zyt}, the effective area for upgoing muon neutrinos reaches approximately \( 1~\text{m}^2 \) at \( 10^4 \)–\( 10^5~\text{GeV} \), depending on zenith angle and event reconstruction. We adopt this value as a representative benchmark to estimate the event rate in a model-independent and conservative manner. This ensures that our conclusion regarding the non-observability of single PBH bursts remains robust and is not reliant on optimistic detector assumptions. This implies that single PBH bursts must occur at much closer distances (e.g., $\lesssim$ a few AU) or be significantly enhanced to yield detectable signals.

\subsection{Event Rate Contours at IceCube} \label{sec:icecube}

The following analysis focuses exclusively on the detectability of an individual PBH burst and should be interpreted as a controlled single-source estimate. A realistic assessment of IceCube sensitivity to PBHs requires modeling a full spatial distribution and abundance, which we address separately in Sec.~\ref{sec:real}.

To further explore detectability at the single-source level, we compute the expected number of muon neutrino events at IceCube as a function of the memory-suppression parameter $k$ and the HNL mass $m_N$, using
\begin{equation} \label{eq:events}
N_{\text{events}} \simeq 
\frac{dN}{dE} \,
A_{\text{eff}} \,
\Delta E \,
(4\pi D^2)^{-1}.
\end{equation}

We consider two benchmark distances for the PBH location. First, we take \( D = 0.01 \) pc, motivated by optimistic local halo scenarios discussed in the literature~\cite{Carr:2016drx, Coogan:2020tuf}. Second, we examine \( D = 1 \) AU as an extreme limiting case that maximizes the geometric flux and therefore provides an absolute upper bound on detectability.

\begin{figure}[h!]
    \centering
    \includegraphics[width=0.8\textwidth]{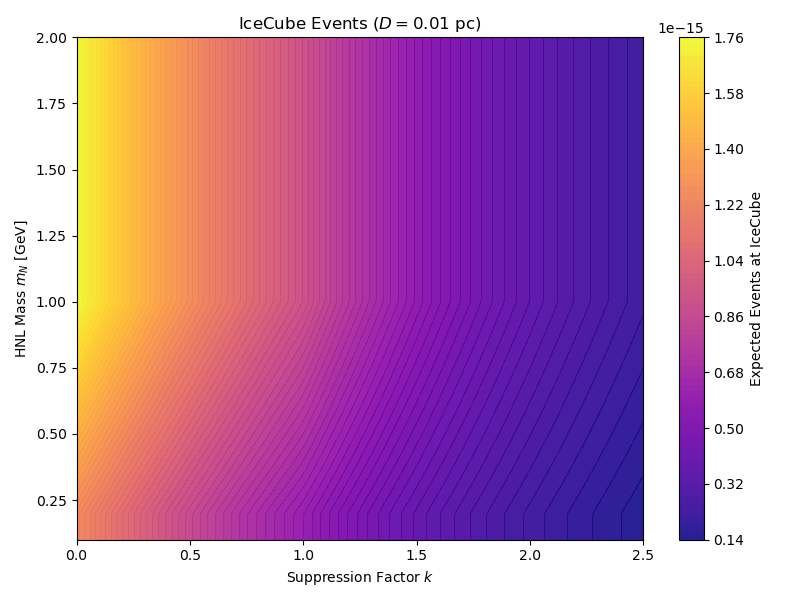}
    \caption{Expected IceCube muon neutrino event count from a PBH at $D = 0.01$ pc, as a function of suppression factor $k$ and HNL mass $m_N$.}
    \label{fig:icecube001pc}
\end{figure}

Figure~\ref{fig:icecube001pc} shows that the event rate increases with decreasing $k$ and increasing $m_N$, reflecting weaker entropy suppression and enhanced HNL decay contributions. Nevertheless, even under these optimistic assumptions, the maximal event rate remains many orders of magnitude below unity. Therefore, a PBH burst located at parsec distances would not be detectable at IceCube.

To explore the absolute upper limit of detectability, we consider a PBH located at 1 AU:

\begin{figure}[h!]
    \centering
    \includegraphics[width=0.8\textwidth]{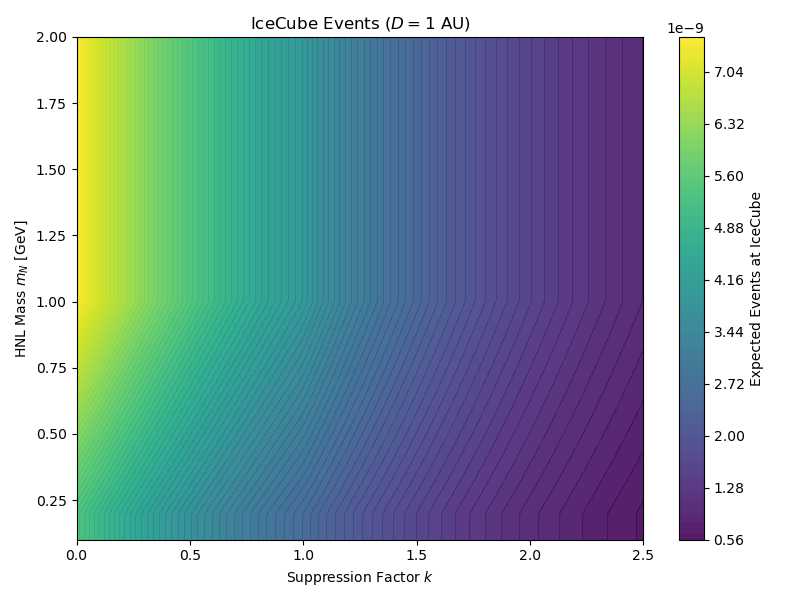}
    \caption{Same as Fig.~\ref{fig:icecube001pc}, but for a PBH at $D = 1$ AU.}
    \label{fig:icecube1AU}
\end{figure}

As shown in Fig.~\ref{fig:icecube1AU}, even at 1 AU — a distance vastly closer than any realistic astrophysical PBH expectation — the predicted event rate remains well below unity across the full $(k, m_N)$ parameter space. This demonstrates that the suppression induced by the memory-burden mechanism, combined with geometric dilution, prevents observable single-source bursts within the considered mass range.

We emphasize that these single-source bounds are complementary to the population-level Galactic estimate presented in Sec.~\ref{sec:real}. Taken together, the two analyses show that neither isolated nearby bursts nor cumulative Galactic contributions yield observable IceCube signals for the parameter space explored.

\subsection{Population-Level Estimate in the Galactic Halo}
\label{sec:real}

The single-source analysis of Sec. \ref{sec:icecube} is illustrative and does not
capture the cumulative contribution from a Galactic population of PBHs.
We therefore estimate the total expected IceCube event count from a
Milky Way halo population assuming that PBHs trace the dark matter (DM)
distribution.

For a PBH mass $M$, the total expected number of detected events from
the Galactic halo is
\begin{equation}
N_{\rm gal}
=
\int dV \,
\frac{\rho_{\rm PBH}(r)}{M}
\frac{N_{\rm burst}(M)}{4\pi D(r)^2},
\label{eq:Ngal_general}
\end{equation}
where $\rho_{\rm PBH}(r)=f_{\rm PBH}\,\rho_{\rm DM}(r)$,
$f_{\rm PBH}$ denotes the fraction of dark matter in PBHs,
$D(r)$ is the distance from Earth,
and $N_{\rm burst}(M)$ is the total neutrino yield per PBH burst folded
with the detector effective area.

Assuming an NFW profile for the Galactic DM halo,
\begin{equation}
\rho_{\rm DM}(r)
=
\frac{\rho_s}{(r/r_s)(1+r/r_s)^2},
\end{equation}
and performing the angular integration analytically, one obtains
\begin{equation}
N_{\rm gal}
=
\frac{f_{\rm PBH}}{2 M R_\odot}
\int_0^{r_{\rm max}} dr \,
r\,\rho_{\rm DM}(r)
\ln\!\left(
\frac{r+R_\odot}{|r-R_\odot|}
\right)
N_{\rm burst}(M),
\label{eq:Ngal_final}
\end{equation}
where $R_\odot\simeq 8.2\,{\rm kpc}$.

\paragraph{Monochromatic case.}
We first consider a monochromatic PBH mass function centered at
$M=10^{10}\,{\rm g}$.
Using the full NFW integral in Eq.~(\ref{eq:Ngal_final}),
and assuming a PBH abundance saturating current cosmological bounds,
$f_{\rm PBH}\sim 10^{-8}$, consistent with existing constraints on evaporating
PBHs in the mass range $M \sim 10^{9}$--$10^{11}\,\mathrm{g}$ from
gamma-ray background and Big Bang Nucleosynthesis observations~\cite{Carr:2020gox,Arbey:2019mbc,Laha:2019ssq},
we obtain
\begin{equation}
N_{\rm gal} \simeq (2\text{--}3)\times 10^{-2}
\quad (k=0).
\end{equation}

Since the expected event count scales linearly with $f_{\rm PBH}$,
this result should be interpreted as conditional on current observational bounds.
An increase of the PBH abundance by one order of magnitude would
proportionally enhance $N_{\rm gal}$; however, such values are already
strongly constrained by existing gamma-ray and cosmological observations.

This result should be contrasted with simplified point-source
scaling estimates in which all PBHs are assumed to lie at a fixed
Galactic-center distance.
The full spatial integration over the NFW halo profile reduces the
effective flux relative to such an approximation and yields a total
event count well below unity even in the absence of memory suppression.
Therefore, population-level stacking within the Milky Way does not
render the signal detectable under standard evaporation assumptions.

\paragraph{Lognormal mass function.}
We next generalize to a lognormal PBH mass distribution,
\begin{equation}
\psi(M)
=
\frac{1}{\sqrt{2\pi}\sigma M}
\exp\!\left[
-\frac{\ln^2(M/M_c)}{2\sigma^2}
\right],
\end{equation}
normalized such that $\int dM\,\psi(M)=1$.
In this case Eq.~(\ref{eq:Ngal_general}) factorizes, and the mass
dependence enters through the average
$\langle N_{\rm burst}(M)/M \rangle$.
Using the approximate scaling
$N_{\rm burst}(M)\propto M^{2-2k}$,
one finds that the enhancement relative to the monochromatic case is
proportional to the mean mass,
$\langle M\rangle = M_c \exp(\sigma^2/2)$.
Even for broad distributions with $\sigma\simeq1$--1.5,
this enhances $N_{\rm gal}$ by at most a factor of a few and does not
change the order-of-magnitude conclusion.
Hence, realistic mass distributions do not qualitatively alter the
detectability estimate.

\paragraph{Impact of memory burden.}
Including entropy-suppressed emission,
$N_{\rm burst}(M)\propto M^{2-2k}$,
further suppresses the Galactic signal.
For modest suppression parameters $k\gtrsim 0.3$,
we find
\begin{equation}
N_{\rm gal} \ll 10^{-2},
\end{equation}
and the expected event count becomes completely negligible.
We therefore conclude that, once a realistic Galactic halo distribution,
observational bounds on $f_{\rm PBH}$,
and memory-burden suppression are consistently taken into account,
the cumulative Milky Way contribution remains far below IceCube
detectability.

We therefore conclude that, once a realistic Galactic halo distribution
and observational bounds on $f_{\rm PBH}$ are taken into account,
population-level stacking within the Milky Way does not yield an
observable signal at IceCube for the parameter space considered,
for PBH abundances consistent with current observational constraints.

\section{Conclusion}

We have investigated the impact of quantum gravitational memory burden effects on the neutrino signals from evaporating primordial black holes (PBHs). By introducing an entropy-suppression parameter \( k \), we quantified how this effect modifies the evaporation dynamics and systematically suppresses the high-energy muon neutrino fluence—especially during the final, high-temperature stages. This suppression imposes severe limitations on the detectability of individual PBH bursts at neutrino observatories such as IceCube.

To assess the influence of beyond-the-Standard-Model physics, we incorporated heavy neutral leptons (HNLs) as additional evaporation products. The subsequent decays of HNLs inject secondary neutrinos, partially mitigating the memory-induced suppression and enhancing the fluence in the GeV range. Nevertheless, even under optimistic conditions—low memory burden, heavier HNLs, and PBHs located within \( D = 0.01 \) pc—the expected number of neutrino events remains well below IceCube’s sensitivity.

Beyond the illustrative single-source estimates, we have performed a population-level calculation of the cumulative Galactic signal by integrating the PBH distribution over a realistic NFW dark matter halo profile. Assuming PBHs trace the dark matter density and saturate current abundance limits, consistent with existing gamma-ray and cosmological constraints on \( f_{\rm PBH} \), we find that the total expected IceCube event count from the Milky Way halo remains well below unity even in the absence of memory suppression.

Generalizing to lognormal mass functions does not qualitatively alter this conclusion, as the enhancement from mass dispersion increases the event rate only by factors of order unity. Including entropy-suppressed emission further reduces the cumulative signal, rendering the Galactic contribution negligible across the viable parameter space.

We emphasize that the expected event rate scales linearly with \( f_{\rm PBH} \); however, values significantly larger than those adopted here are already disfavored by current observational bounds.

We therefore conclude that, once realistic spatial distributions and observational bounds on \( f_{\rm PBH} \) are consistently taken into account, and for PBH abundances compatible with current observational constraints, neither individual bursts nor Galactic population-level stacking yield observable signals at current IceCube sensitivity for the mass range considered.

Our findings highlight the interplay between quantum gravity corrections and exotic particle content in shaping PBH evaporation signatures. While present neutrino telescopes are unlikely to detect memory-burdened PBH evaporation in this parameter regime, the framework developed here provides a systematic approach to incorporating quantum gravitational suppression effects into multi-messenger PBH searches. Future improvements in detector sensitivity or alternative observational channels may further test these scenarios.

\section{Acknowledgment}
The work of AC was partly supported by the Japan Society for the Promotion of Science (JSPS) as a part of the JSPS Postdoctoral Program (Standard), grant number: JP23KF0289. KP acknowledges the University Grants Commission(UGC) for supporting his doctoral work. The authors gratefully acknowledge Banaras Hindu University for hosting the XXVI DAE-BRNS High Energy Physics Symposium 2024, where this work was initiated.

\end{document}